# Stalled phase transition model of high-elastic polymer

Vadim V. Atrazhev[1,2,*], Sergei F. Burlatsky[3], Dmitry V. Dmitriev[1,2], Vadim I. Sultanov[2]

[1] Russian Academy of Science, Institute of Biochemical Physics, Kosygin str. 4, Moscow, 119334, Russia

[2] Science for Technology LLC, Leninskiy pr-t 95, 119313, Moscow, Russia

[3] United Technologies Research Center, 411 Silver Lane, East Hartford, CT 06108, USA

[*] Corresponding author. Tel.: +7 495 939 08 88; fax: +7 499 137 82 31.

*E-mail address:* vvatrazhev@deom.chph.ras.ru (V.V. Atrazhev)

**Abstract**

The microscopic model of semi-crystalline polymer in high-elastic state is proposed. The model is based on the assumption that, below the melting temperature, the semi-crystalline polymer comprises crystal nuclei connected by stretched chain segments (SCS) with random configuration of monomers. The key factor that stalls the phase transition below the melting temperature is the tension of the SCS. External stress applied to the polymer also shifts the equilibrium and causes unfolding of the nuclei, which enables large reversible deformation of the polymer without loss of integrity. The simple 1D model predicts plateau in stress-strain curve of high-elastic polymer above the yield stress, which agrees with experimental observations. The model prediction for the temperature dependence of polytetrafluoroethylene (PTFE) yield stress in high-elastic state is in satisfactory agreement with experiment.

**1. Introduction**

The modeling of mechanical properties of solid polymers is important for multiple industrial applications. The popularity of polymer materials is caused by a large variability of their mechanical properties. For example, elastic modulus of solid polymers varies from several MPa for rubbers to several GPa for polyamides. Design of new polymers and polymer composite materials could be accelerated by the model, which relates the polymer structure and composition to the mechanical properties. The microscopic model of polymer mechanical properties also can be helpful for prediction of polymer components durability, e.g. durability of polymer membranes [1,2]. Polymers durability under mechanical cycling conditions is governed by kinetics of chain free radical reactions [3]. The kinetics of chain reactions in polymer systems depends on the microscopic structure of the polymer and microscopic stress distribution [3].

Thereby, the polymer durability modeling can be approached using the theory of chemical reactions in polymer systems [4] and the microscopic inter-chain stress distribution calculated by the microscopic model of polymer mechanical properties. In this paper, we propose the microscopic model of deformation of semi-crystalline polymer with the flexible chains that was developed based on the hypothesized microscopic polymer structure.

Polymer mechanical properties strongly depend on the temperature. Typically, three aggregate states of the polymer are referred in the literature: polymer melt, rubber or entropy-elastic state, energy-elastic or glassy state [5]. The later two states are considered as the solid states of the polymer. Rubber is also referred to in the literature as a high-elastic state. Temperature transition from the melt to the solid polymer is a first order phase transition with the heat release [6]. Moreover, melting of semi-crystalline polymers is a complex phase transition due to the metastability of their structure and a number of semi-crystalline polymers exhibit double or multiple melting upon heating in differential scanning calorimetry [7].

Above the melting temperature, the polymer behaves as a very viscous liquid. Below the melting temperature, the polymer conserves the shape as a solid. The major differentiator of the solid polymers from crystalline matter and low-molecular glass is that the elastic modulus of the polymer is by two orders of magnitude smaller and that the polymer elongation to break is extremely large (> 100%) [8,9]. The temperature transition from the high-elastic state to the glassy state is a relaxation transition in which the relaxation time of the polymer sharply changes. Mechanical properties of the polymer change drastically at the glass transition temperature, $T_g$. However, this transition occurs not at fixed temperature as phase transitions do but in a narrow temperature region near $T_g$ [8]. The nature of glass transition is not clear so far. The current understanding of the glass transition phenomenon can be found in the review [10]. The behavior and properties of polymer glasses substantially differ from that of the low-molecular glasses. Polymers typically demonstrate at least two glass transition temperatures, which are detected by dynamic mechanical analysis (DMA) test and referred to in the literature as α-relaxation and γ-relaxation [11,12]. Above the α-relaxation temperature, most polymers demonstrate high-elastic properties: low elastic modulus and high (>100%) reversible elongation. Elastic modulus of the high-elastic polytetrafluoroethylene (PTFE) increases with the increase of the temperature [12]. Such temperature dependence of elastic modulus is specific for entropic elasticity of cross-linked rubbers. At α-relaxation temperature, $T_\alpha$, the elastic modulus changes by the order of magnitude. Below $T_\alpha$, elastic modulus increases with decrease of temperature. In addition, only a small fraction of deformation (~5%) remains reversible. The large deformations become irreversible, *i.e.* a plastic deformation, and the polymer does not recover its shape after unloading. However, the large deformations produced below $T_\alpha$ can be recovered by the heating of the polymer to the temperatures higher than $T_\alpha$. This phenomenon is known as thermally stimulated recovery [13–16]. The polymer loses plasticity below γ-relaxation temperature, i.e. relatively small elongation causes the polymer brittle fracture. Fundamental bases of modern understanding of polymer melt were founded by Flory [17], Huggins [18,19], Doi and Edwards [20] and de Gennes [21,22]. Flory-Huggins theory describes the thermodynamics of the polymer solution. The dynamics of polymer melts and/or concentrated polymer solutions is described by well-known 'reptation' mechanism proposed by de Gennes [21]. The reptation model explains the experimentally observed dependence of the macromolecule mobility on the molecular weight and is used as a mechanism to explain the viscous flow in an amorphous polymer. The coil-globule transition model was studied in pioneer work of Flory [23], Lifshits, Grosberg and Khokhlov [24]. The theory of coil-globule transition in semi-dilute solutions was further developed in [25,26] where the new mechanism of coil-globule phase transition was discovered. This mechanism involves subsequent formation of polymer folds on different spatial scales and results in formation of self-similar, fractal, globular structures and peculiar kinetic law. The fundamental introduction into statistical physics of polymer solution and polymer melt can be found in [27] and [28].

The mechanical model of vulcanized rubber was the first model of the solid comprised of macromolecules. The phenomenon of rubber elasticity can be understood qualitatively and, to a large extent, quantitatively in terms of statistical theory developed by James and Guth [29] and Flory [30] and Treloar [31]. The model treats the vulcanized rubber as a network of cross-linked polymer chains, and a thermodynamic equilibrium in the system is assumed. The entropy of the chain segments depends on the distance between cross-links. At the polymer deformation, the distance between the cross-links changes and the entropy of the network decreases. As a result, the free energy increases, which manifests itself in the elastic force that acts against the deformation. The feature of the entropic elasticity is a small value of elastic modulus, equal to approximately a few MPa. The rubber elastic modulus is proportional to the absolute temperature and inversely proportional to the concentration of the cross-links, which is in a good agreement with experimental data for the cross-linked rubbers [30]. The further modifications of the rubber elasticity model were proposed in the later works [32–37].

Mechanical properties of semi-crystalline polymers in high-elastic state differ from that of cross-linked rubbers. Elastic modulus of high-elastic PTFE (above $T_\alpha \sim 130°C$) increases with increase of the temperature, which agrees with prediction of the rubber elasticity model. However, the value of the elastic modulus of the high-elastic polymer can be much larger than that of the cross-linked rubber. For example, the elastic modulus of the PTFE above $T_\alpha$ is approximately equal to 50 MPa [38]. The stress-strain curves of high-elastic semi-crystalline polymer demonstrate the abrupt increase of the stress at low deformations (<30%) followed by the region with weak dependence of the stress on the strain [39] while the strain of the cross-linked rubber linearly increases with increase of the stress [29], as shown in Fig. 1. Thus, the experimental data suggest that the rubber elasticity model is not directly applicable to the semi-crystalline polymer in the high-elastic state.

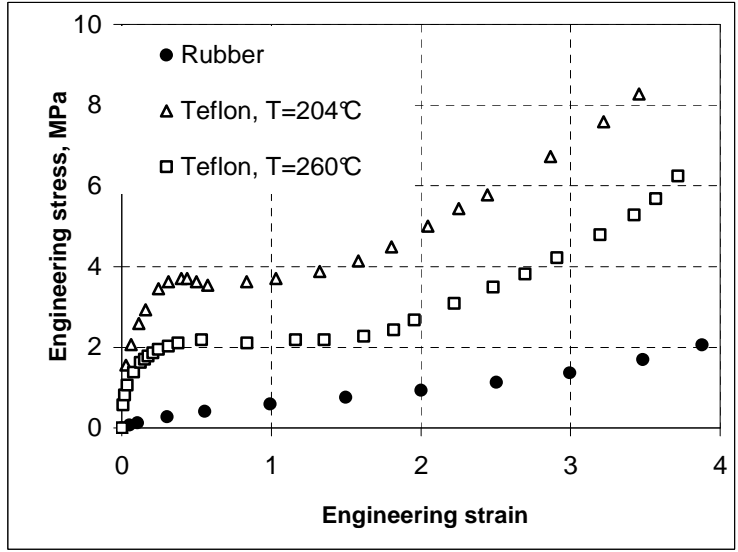

**Fig. 1.** Stress-strain curves of rubber at room temperature [29] and PTFE in high-elastic state at T=204°C and T=260°C [39].

Haward proposed the phenomenological model of deformation of the glassy polymer [40,41]. The model utilizes the analogy with mechanical system consisting of Hookean spring in series with a dashpot, which is in parallel with another spring. The microscopic interpretation of the Hookean springs is a polymer Gaussian coil, which provides the rubber-like entropic elasticity. The dashpot results in the force of viscous friction, which depends on the deformation rate. The model predicts the Gaussian equation for true stress, $\sigma_{true}$, vs. polymer elongation, $\lambda$:

$$\sigma_{true} = Y + G_p\left(\lambda^2 - \frac{1}{\lambda}\right)$$

Here $Y$ is extrapolated yield stress; $G_p$ is a strain hardening modulus. Many experimental data for glassy polymers are in a good agreement with Gaussian equation [40]. One of the model limitations is related to the temperature dependence of $G_p$. The rubber elasticity model predicts increase of $G_p$ with increase of the temperature, while experimental dependence of $G_p$ for the glassy polymers demonstrates a reverse trend. The further modification of the model includes dependence of $Y$ on deformation rate using the Eyring equation [42], which predicts that the deformation rate exponentially depends on the applied stress [43]. The Gauss-Eyring model predicts logarithmical dependence of $Y$ on deformation rate, which agrees with the most of the experimental data for the glassy polymers. The Eyring model was extended to the case of stress relaxation under constant loading in [2]. However, the microscopic interpretation of the yield stress of high-elastic polymers is still unclear.

The modern microscopic models of the solid polymers take into account semi-crystalline molecular structure of the polymers below the melting point. Semi-crystalline polymers consist of both amorphous and crystalline domains, where the percentage of crystallinity can vary from 10% to 90% in commercially available materials [8,9,44]. The polymer elastic properties are attributed to the amorphous regions, which are considered as a solid melt of the randomly packed polymer chains. The cluster model [45,46] of the solid polymer was developed during the last three decades. According to the cluster model, the structure of the amorphous regions represents a network of clusters connected by the passing chains and inter-cluster matrix comprised of the randomly packed and entangled chains. The clusters are small regions of local ordering consisting of the regular packed chain segments of the different macromolecules. The cluster model successfully explains many experimental observations. However, it contains a large number of phenomenological parameters, which reduces the predictive power of the model.

In this paper, we propose the microscopic model of deformation of the high-elastic polymer. The model has a similarity with the cluster model. We also consider the amorphous regions of the polymer as the network of clusters connected by the chain segments. However, in our model the repeating unit of the network is crystalline nucleus that can exchange monomers with the stretched chain segment (SCS) that connect the nuclei. A similar structure of the polymer below the melting point was observed in molecular dynamics (MD) simulations of the polymer crystallization process [47–50]. Muthukumar proposed the model of polymer crystallization in which the crystallization process is divided into three stages. The first stage is formation at the chain of several "baby nuclei", connected by the same single chain. The strands connecting these baby nuclei are flexible with considerable configurational entropy. As time progresses, the monomers in the flexible strands are reeled into the baby nuclei while the orientational order in each nucleus increases making them "smectic pearls" [49]. At the third stage, the smectic pearls attach to each other and form lamellar structure. We assume that in the amorphous part of the polymer, the attachment of the nuclei is stalled by topological restrictions and major number of nuclei does not grow into the large crystals far below melting point.

In the present model, the mechanical properties of the high-elastic polymers are governed by thermodynamics of the nucleus/SCS monomer exchange. The model analytically calculates the yield stress of the high-elastic polymer as a function of the temperature. In the model, the large deformation of the polymer is enabled by the transfer of the monomers from the nuclei to the SCS. That agrees with recent MD study of mechanical properties of semi-crystalline polyethylene [51]. The basic mechanism of plastic deformation observed in [51] is the pull-out of the chains from the crystalline fraction of the polymer. Conceptually, our model of the polymer deformation is very similar. However, the yield stress ~0.5-1 GPa obtained in [51] is much higher than the experimental one. We speculate that this is a consequence of a short time interval available in the MD simulations. Our model predicts analytically the yield stress value that is close to experimental one for PTFE and reproduces the experimentally observed increase of the yield stress with decrease of the temperature.

The physics and major features of the analytical expression for the tensile stress of the SCS predicted by the current model are very similar to the results obtained by Di Marzio and Guttman [52] for the peeling force of a single polymer chain absorbed on the surface. In both cases, the equilibrium between the low and high energy states of the chain monomers governs the force value. For high-elastic deformation, this is the equilibrium between the nucleus and the SCS. For chain peeling, this is the equilibrium between the absorbed and peeled out fractions of the chain. Consequently, both the peeling force and the tensile force of the SCS have similar qualitative features.

The paper is organized as follows. The hypothesized microscopic structure of high-elastic polymer is presented in section II. The thermodynamic model of the polymer deformation that calculates the apparent yield stress of the high-elastic polymer is presented in section III. Comparison of the model predictions for high-elastic PTFE yield stress with experimental data is also presented in this section. The molecular dynamic simulation of the polyethylene nucleus formation and melting is presented in section IV. The model results are summarized and discussed in section V.

## 2. Microscopic structure of solid polymer

In this section, we propose the model of microscopic structure of the semi-crystalline polymer below the melting temperature, $T_0$. In low-molecular liquids, crystallization process starts from formation of nuclei, develops through the stage of the nuclei growth and finishes by formation of monocrystal or polycrystalline granular structure. This phase transition is completed at a constant temperature below $T_0$ in a single component system. According to [48,53], crystallization in the polymer melt also starts with nucleation process. However, topological restrictions in polymer melt [54] caused by the connectivity of monomers into the chain impede the growth stage. As a result, the major number of nuclei does not grow into the large crystals far below $T_0$.

The chains in the polymer melt are randomly packed and entangled as shown in Fig 2a. The high level of entanglement restricts the conformational motion of the chains in the melt. Particularly, this restrictions result in switching from Rouse chain dynamics [20,27,55] in diluted polymer solutions to reptation in the polymer melt [21,22]. We assume that the reptation mechanism is suppressed after the start of nucleation of the melt below the melting point.

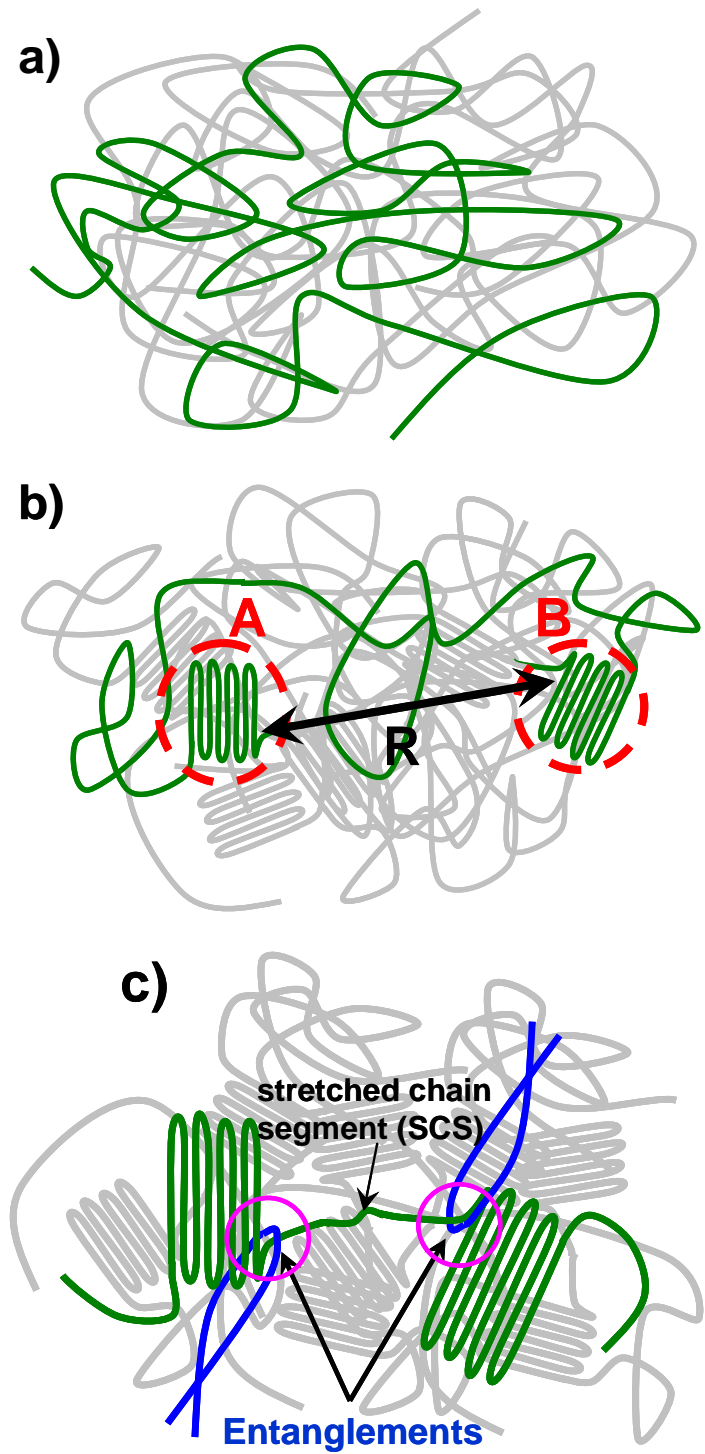

**Fig. 2.** The evolution of polymer structure at fixed temperature below $T_0$. a) polymer melt, b) beginning of nucleation process (nuclei A and B are the neighboring nuclei belonging to one chain), c) termination of nuclei growth (nuclei A and B are arrested by entanglements), the SCS is stretched between the nuclei A and B.

We hypothesize that at fixed temperature below $T_0$ a large number of crystal nuclei forms at each chain. In this paper, we consider the flexible polymer molecules such as PTFE or polyethylene (PE) that nucleate by forming folds, as shown in Fig. 2(b). This structure of the nuclei agrees with experimental observations of the structure of PTFE and PE crystals, with our MD simulation of PE crystal presented in section IV and with MD simulation results presented in [47,48,50]. Some nuclei comprises of the monomers of single chain. Other nuclei comprises of the monomers of two or more chains. The stretched chain segment (SCS) connects two neighboring nuclei belonging to one chain (A and B in Fig. 2(b)). These nuclei can be separated in the space by other nuclei, as shown in Fig. 2(b). Below $T_0$, the nuclei A and B grow consuming the monomers of the SCS. The reduction of the number of monomers in the SCS generates a tension force, *f*, of the SCS, which pulls the nuclei A and B towards each other. That can result in formation of agglomerates of several nuclei and higher scale structures such as lamellas and spherulites [56,57]. However, the excluded volume effects and topological entanglements balance this tension force preventing a fraction of nuclei from collapsing. We speculate that reptation of the chain cannot pass through the crystal nuclei. We assume that the typical radius of the curvature of the entanglements loop in polymer melt is of the order of the polymer persistent length (~1nm). The typical size of the nuclei at the initial stages of nucleation is of the order of several nanometers [47,49]. At the end of crystallization process the lamellar structure of the semi-crystalline polymer is formed. The thickness of lamella is about 10-50 nm and breadths can be even larger [58]. Therefore, the nucleus cannot penetrate through entanglement loop even at initial stage of nucleation, especially at the stage of formation of lamellar structure. Consequently, the nuclei motion is arrested at some distance *R* between the nuclei, as shown in Fig. 2(c). The result of crystallization process described above is a network of the nuclei connected by the SCS.

We hypothesize, that the dynamic equilibrium between the nuclei and the SCS is established in the area, indicated by pink circles in Fig. 2(c). The thermodynamic driving force pushes the monomers from the SCS into the nucleus. The SCS tension force pulls the monomers from the fold into the SCS. The nuclei tend to consume the monomers of the SCS, which results in tension of the SCS. This tension pulls the nuclei A and B toward each other. The excluded volume interaction balances the tension. As a result, the nuclei network is formed. This network is characterized by the numbers of monomers in the nuclei and in the SCS and by the distances between the nuclei. These parameters depend on the nature of the polymer and on preparation procedure. The tension force of the SCS results in the internal stress in the polymer.

The number of the monomers in the SCS is minimal at zero absolute temperature, *i.e.* the free segments are fully stretched, as shown in Fig. 3(a). At non-zero temperature, some fraction of the monomers transfers from the nuclei to the SCS to increase the entropy, as shown in Fig. 3(b). When the temperature approaches $T_0$, all the monomers transfer from the nuclei into the SCS. The size and volume fraction of the nuclei approaches zero, as shown in Fig. 3(c).

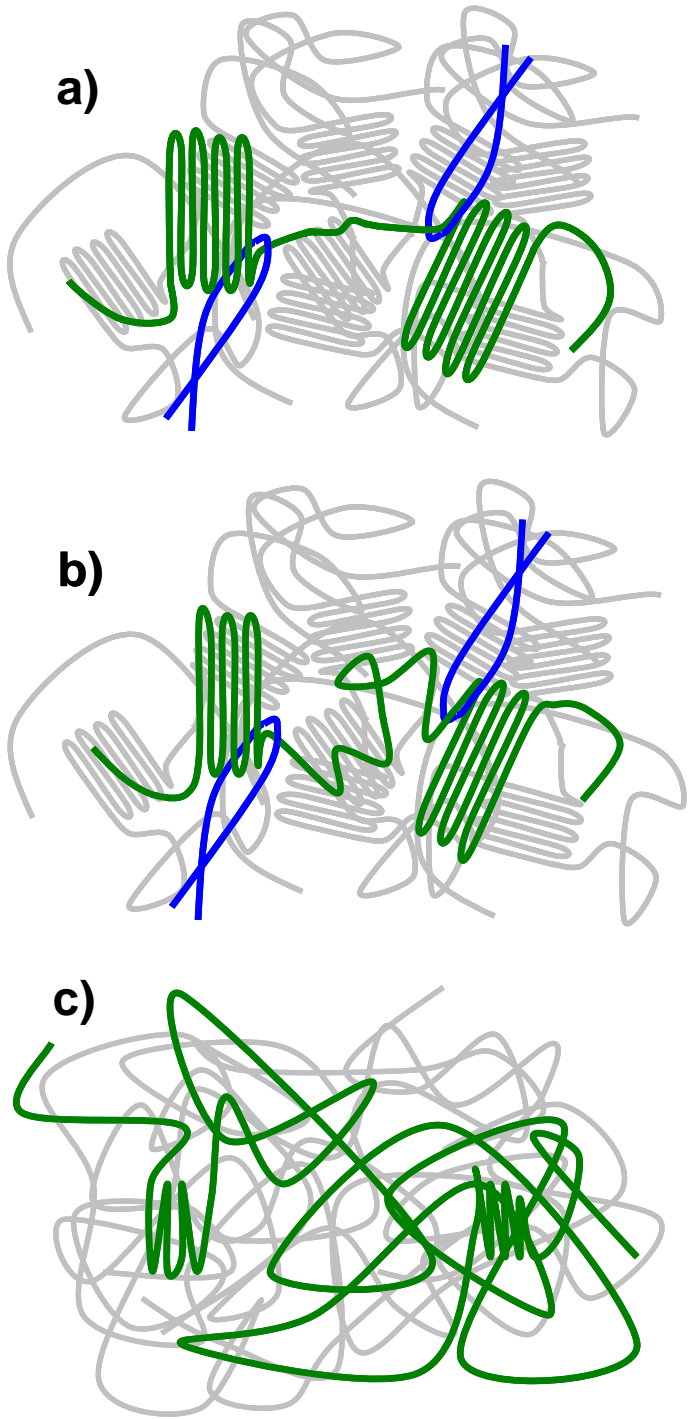

**Fig. 3.** The polymer structure at different temperatures a) $T = 0$, the number of the monomers in the SCS is minimal and the SCS is fully stretched, b) $T > 0$, some fraction of the monomers transfers from the nuclei into the SCS, c) $T \rightarrow T_0$ (melting temperature), the major fraction of the monomers transfers into the SCS.

An external stress applied to the polymer shifts the thermodynamic equilibrium between the nuclei and the SCS. In equilibrium, the forces acting on the nucleus from the neighboring nuclei balance the tensile force of the SCS. The external stress disturbs this balance and causes relative motion of the nuclei. Large elongation of the polymer in *x*-direction leads to the transfer of the monomers from the nuclei A and B to the SCS oriented along *x*-direction, as shown in Fig. 4. This process can be considered as the melting of the nuclei induced by the external stress. To conserve the polymer density the other nuclei move in *y*- and *z*-directions towards each other and fill the free volume emerged after the extension of the nuclei A and B. The initial shape of polymer can recover after unloading. The distance *R* and the number of monomers in the nuclei and in the SCS also can return to initial values after unloading. This process can be considered as reverse crystallization of the nuclei after unloading. The melting/crystallization of the nuclei under external stress enables large reversible deformation of the polymer above the glass transition temperature.

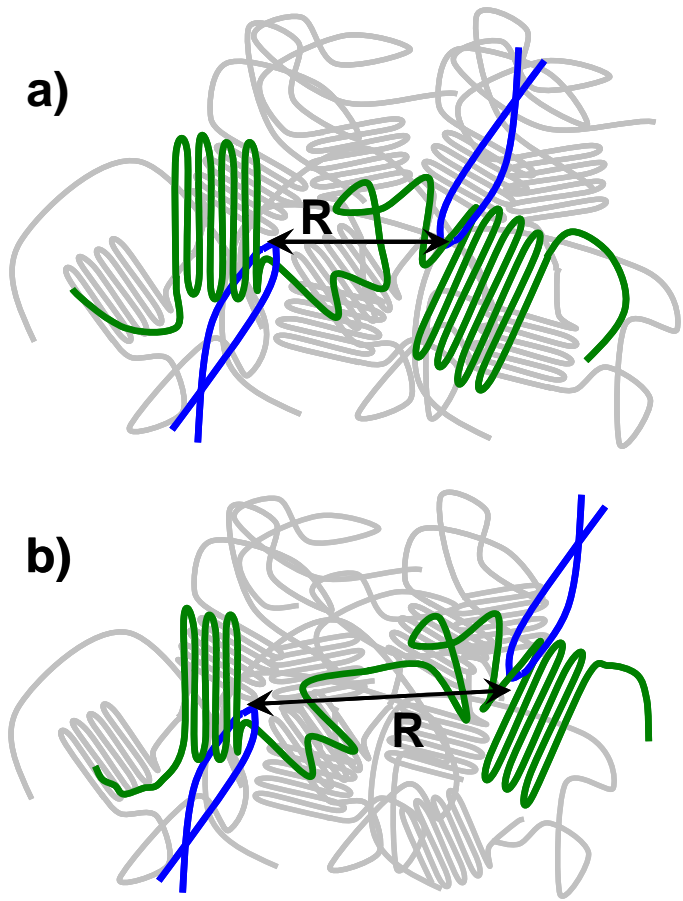

**Fig. 4.** The unfolding of the nuclei induced by large deformation. The system a) before deformation, b) after deformation. The distance *R* and the number of the monomers in the SCS after deformation are larger than that before the deformation. The number of the monomers in the nuclei is smaller after deformation than that before the deformation.

In this section, we proposed microscopic model of the semi-crystalline polymer, which considers the polymer as the network of the crystal nuclei connected by the randomly packed stretched chain segments (SCS). The monomers in each SCS are in dynamic equilibrium with the monomers in the nuclei. This dynamic equilibrium is governed by the monomer to monomer attraction force, temperature and the tension of the SCS that depends on the applied external stress and the temperature.

## 3. Thermodynamic model of polymer deformation

In this section, we present a simplified one-dimensional thermodynamic analysis of the model that was discussed in section II. According to this model, the number of the monomers in the SCS is governed by the competition of the tendency of the nucleus to grow to decrease the energy and the tendency to increase the number of monomers in the SCS to increase the entropy and decrease the tension force. Below we calculate the tensile force of the SCS as a function of temperature and estimate the yield stress of the polymer above the glass transition temperature. We start with the Gaussian approximation for the SCS entropy and neglect the surface energy of the nuclei. Then we discuss more general non-Gaussian distribution using scaling arguments and incorporate the surface energy into the model. Finally, we qualitatively compare the model predictions for the yield stress to available experimental data for PTFE.

As described in Sec.2 we model the polymer as a nuclei network, where small crystalline nuclei are separated by amorphous regions and/or other nuclei. The nucleus with the SCS constitutes a repeating element of the polymer network. Our model of polymer as nuclei network is characterized by two parameters: the numbers of monomers in the repeated unit (nucleus plus the SCS), $N$, and by the distances between the nuclei, $R$. The fraction of the numbers of the monomers in the SCS, $n$, and nucleus, $(N-n)$, is not fixed, it is governed by thermodynamics and, therefore, depend on the temperature.

We start with the simplest model that does not take into account the surface energy of the nucleus. The free energy of the repeating element, $F(n, R)$, is:

$$F(n,R) = n(E_0 - TS(n,R)) \tag{1}$$

Here $E_0$ is the difference of the monomer energy in the nucleus and in the SCS. The $E_0$ is equal to the specific heat of melting per one monomer. $S(n, R)$ is the conformational entropy of the SCS per one monomer. The conformational entropy of the nucleus is equal to zero.

The conformations of the SCS near $T_0$ are similar to the conformations of the free chain segment between two entanglements. According to Flory theory [17], the entropy of the SCS near $T_0$ is approximately equal to the entropy of the free chain segment containing $n$ monomers and fixed between two entanglements in the points A and B. Here we model the SCS as a random walk containing $n$ steps, started from the point A and finished in the point B. We assume that $na \gg R$ and Gaussian approximation for random walk is applicable. The entropy of the Gaussian coil with the distance $R$ between the ends of the chain is:

$$S_G = nk \ln(z) - k \frac{3R^2}{2na^2} \quad (2)$$

Here $z$ is the number of states per one monomer in the melt, $a$ is the monomer length and $k$ is Boltzmann constant. The parameter $ln(z)$ can be calculated through $E_0$ and $T_0$ as follows. The entropy of the melt is

$$S_m = nk \ln z \quad (3)$$

At the melting point, the free energy of the melt is equal to the free energy of the crystal

$$nkT_0 \ln z = nE_0 \quad (4)$$

Solving equation (4) with respect to $ln(z)$ we obtain

$$\ln(z) = \frac{E_0}{kT_0} \quad (5)$$

Substituting equations (2) and (5) into (1) we obtain the equation for the free energy of the repeating unit:

$$F(n,R) = nE_0 \frac{T_0 - T}{T_0} + kT \frac{3R^2}{2na^2} \quad (6)$$

The majority of polymer materials are semi-crystalline, though the thermodynamically stable state is monocrystale [58]. The semi-crystalline polymers are in thermodynamically metastable states that are separated by multiple high free energy barriers [45,53]. Transition of the polymer into the global minimum is hindered by topological entanglements and excluded volume effects and can take longer than the polymer lifetime. Polycrystalline polymer consists of large number of crystals separated by amorphous fraction. Global optimization of equation (6) would result in trivial solution $R=0$. That corresponds to the system with zero distance between nuclei, *i.e.* polymer monocrystal. In this paper, we constrain the system postulating the mean distance between the nuclei, $R$, at the local minimum and analyze the mechanical properties of the polymer as a function of $R$ during elongation process.

We assume thermodynamic equilibrium between the nucleus and the SCS. That implies fast exchange of the monomers between the nuclei and the SCS that results in the minimization of the free energy with respect to the number of the monomers in the SCS during the deformation or temperature change. The number of the monomers in the SCS at fixed distance $R$ is calculated through minimization of the free energy:

$$\frac{\partial F(R,n)}{\partial n} = 0 \qquad (7)$$

Substituting equation (6) for the free energy into (7) we obtain

$$E_0 \frac{T_0 - T}{T_0} - kT \frac{3}{2}\left(\frac{R}{na}\right)^2 = 0 \qquad (8)$$

Solving equation (8) with respect to $n$ we obtain the equilibrium number of the monomers in the SCS

$$n_{eq}(T,R) = \frac{R}{a}\sqrt{\frac{3kT_0}{2E_0}\frac{T}{(T_0 - T)}} \qquad (9)$$

The equilibrium number of the monomers in the SCS is a function of temperature and the distance between the neighboring nuclei, as follows from equation (9). Equation (9) predicts that at temperature $T_0$ the number of the monomers in the SCS tends to infinity. The singularity at the phase transition point is a typical feature of many phase transition models. Typically, this divergence transforms to finite spike when finite size effects are taken into account. The divergence of the number of the monomers in equation (9) at $T_0$ was eliminated when the non-leading logarithmic term was incorporated into the expression for the entropy of Gaussian (see Appendix). At the temperatures below $T_0$ this term is small, however, it changes the system behavior in the vicinity of $T_0$.

Substituting equation (9) into (6) we obtain the equation for the free energy as a function of $T$ and $R$

$$F_{eq}(T,R) = E_0 \frac{R}{a}\sqrt{6\frac{kT}{E_0}\frac{T_0 - T}{T_0}} \qquad (10)$$

The free energy of the SCS linearly depends on $R$, as follows from equation (10). The tensile force that attracts the neighboring nuclei to each other is equal to the derivative of the free energy with respect to the distance between the nuclei, $R$. Differentiating equation (10) with respect to $R$ we obtain the equation for the tensile force, $f_{eq}$, acting on the SCS:

$$f_{eq}(T) = \frac{E_0}{a}\sqrt{6\frac{kT}{E_0}\left(\frac{T_0 - T}{T_0}\right)} \qquad (11)$$

Tensile force of the SCS vanishes when the temperature approaches $T_0$. The tensile force increases as $\sqrt{\Delta T}$ ($\Delta T = T_0 - T$) when the temperature decreases. Such temperature dependence of the tensile force is a consequence of the Gaussian approximation for the conformational entropy of the SCS. The tensile force (11) results in internal stress in the polymer below melting point, which is caused by the tendency of the polymer to crystallize. Equation (11) is incorrect for very low temperatures. Since our assumption about thermodynamic equilibrium between the nuclei and the SCS is wrong below $T_g$, the low-temperature is not relevant.

Equation (2) for the entropy of the SCS is valid near the melting point, where the Gaussian approximation for the SCS conformations is applicable. Thus, equations (9) and (11) are also valid near the melting point. The length of the SCS, *na*, approaches the distance R and the Gaussian approximation fails when the temperature becomes substantially lower than $T_0$. Moreover, far below the melting point the essential fraction of the volume is occupied by the nuclei and is inaccessible for the SCS. That also results in deviation of the statistics of the SCS from the Gaussian statistics. The detailed model of the SCS entropy far below the melting point is out of the scope of current paper. Below we demonstrate that the application of a non-Gaussian approximation for the entropy of the SCS does not break the conclusion about the independence of the tensile force of the SCS on the distance R.

We generalize the conclusion about the independence of the tensile force of the SCS on the distance *R* relaxing the Gaussian approximation for the entropy of the SCS. We use more general expression for the entropy of the SCS, assuming that the entropy of the SCS is a function of a lumped parameter $x = \frac{na}{R}$. Without loss of generality, we write the entropy of the SCS as

$$S_{SCS} = n(k \ln z - \Delta S(x)) \qquad (12)$$

The arbitrary function $\Delta S(x)$ is the loss of the entropy per one monomer in the SCS with respect to the chain segment in the melt. Substituting equation (12) and (5) into (1) we obtain

$$F(n,x) = n\left( E_0 \frac{T_0 - T}{T_0} + T \Delta S(x) \right) \qquad (13)$$

Substituting equation (13) into (7) and taking advantage of $n = x\frac{R}{a}$ we obtain following equation for the optimal number of the monomers in the SCS:

$$E_0 \frac{T_0 - T}{T_0} + T \frac{d(x \cdot \Delta S(x))}{dx} = 0 \qquad (14)$$

Equation (14) includes the number of the monomers in the SCS, *n*, only in the dimensionless combination $x = \frac{na}{R}$. Therefore, the solution of equation (14), non-dimensional equilibrium number of the monomers in the SCS, $x_{eq}(T)$, is only a function of the temperature, i.e. $x_{eq}(T)$ is independent of *R*. The equilibrium number of the monomers in the SCS is

$$n_{eq}(T,R) = \frac{R}{a} x_{eq}(T) \qquad (15)$$

The specific function $x_{eq}(T)$ depends on the function $\Delta S(x)$. Substituting $x_{eq}(T)$ into equation (13) for free energy, we obtain the average energy of the repeating unit as a function of $T$ and $R$:

$$F_{eq}(R,T) = \frac{R}{a} x_{eq}(T) \left( E_0 \left(1 - \frac{T}{T_0}\right) + T\Delta S(x_{eq}(T)) \right) \tag{16}$$

Differentiating equation (16) with respect to $R$ we obtain the equation for the tensile force of the SCS:

$$f_{eq}(T) = \frac{x_{eq}(T)}{a} \left( E_0 \left(1 - \frac{T}{T_0}\right) + T\Delta S(x_{eq}(T)) \right) \tag{17}$$

The tensile force depends only on the temperature and does not depend on the distance $R$ between the nuclei under assumption that the entropy of the SCS depends only on the scaling parameter $x = \frac{na}{R}$.

The external force should be applied to the nuclei to move apart two neighboring nuclei connected by the SCS. This external force acts against the SCS tensile force. The external force that is smaller than the tensile force is not sufficient to initiate the nuclei displacement. The nuclei displacement starts when the external force exceeds the internal tensile force of the SCS. Therefore, the critical stress is required to initiate the large non-Hookean deformation of the polymer. This critical value of the stress manifests itself as a yield stress in the polymer stress-strain curve of the high-elastic polymer. In addition, equation (11) and equation (17) predict that the tensile force is independent of the distance $R$ between the connected nuclei, i.e. the stress in the high-elastic polymer is independent of deformation. Experimental stress-strain curve of PTFE demonstrates the plateau from $\varepsilon=30\%$ to $\varepsilon=200\%$, as shown in Fig. 1. We speculate that this plateau is a result of the internal tensile stress in the polymer that is caused by the SCS tensile force. The origin of the gradual stress increase at large deformations ($\varepsilon > 200\%$) is not discussed in this work. We believe that at the Hookean linear stage of stress-strain curve the polymer deforms without unfolding of the nuclei. However, both, the mechanism of small Hookean deformation and the origin of the gradual stress increase at large deformations ($\varepsilon > 200\%$) are out of the scope of the current work.

In the model presented above, we did not take into account the surface energy of nuclei. However, the effect of the surface energy is important near the melting temperature. We incorporate surface energy terms that were derived in [59] into equation (6):

$$F(n,R) = n\left( E_0 \frac{T_0 - T}{T_0} + T\Delta S(n,R) \right) + \sigma_1 2L^2 + \sigma_2 4hL \tag{20}$$

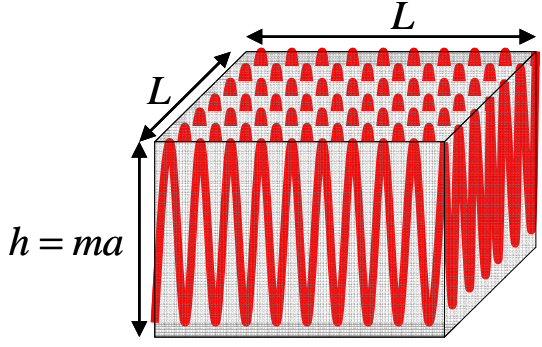

**Fig. 5.** The sketch of folded nucleus containing *(N-n)* monomers.

The sketch of the nucleus is presented in Fig.5. The nucleus of the size $L{\times}L{\times}h$ contains ($N$-$n$) monomers of polymer. In the nuclei the polymer chain is folded along the vertical direction. The size $h$ is fixed and size $L$ varies during melting/crystallization. The quantities $\sigma_1$ and $\sigma_2$ in Eq.(20) are the surface tensions of the nuclei/melt interface for two surfaces $L{\times}L$ and four surfaces $L{\times}h$, respectively.

The mass balance relates the size of the nucleus $L$ with the number of the monomers in the nucleus, $N$-$n$, and the nucleus thickness, $ma$:

$$L = \left(\frac{N-n}{m}\right)^{1/2} a \qquad (21)$$

Substituting equation (21) into (20) we obtain:

$$F(n,R) = nE_0 \frac{T_0 - T}{T_0} + kT \frac{3R^2}{2na^2} + \frac{2E_1}{m}(N-n) + 4E_2\sqrt{m(N-n)} \qquad (22)$$

Here $E_1 = \sigma_1 a^2$ and $E_2 = \sigma_2 a^2$ are the surface energies of the nucleus per one monomer.

The typical dependence of the free energy (22) on *n* for several temperatures is presented in Fig.6. At high temperature (560 K) the minimum of the free energy is archived at *n=N*, which corresponds to the state of all monomers in the melt. At lower temperature (540 K) a new local minimum is obtained at low values of *n*. For this temperature, this local minimum corresponds to a metastable state because the global minimum is still in the melt phase (at *n=N*). When the temperature is further decreased, the local minimum becomes more pronounced (520 K). At some temperature the free energy in this minimum becomes equal to the free energy of the melt. At lower temperatures, the state with small *n* is a global minimum and melt transforms into the nuclei (500 K), i.e. the first-order transition occurs at this temperature.

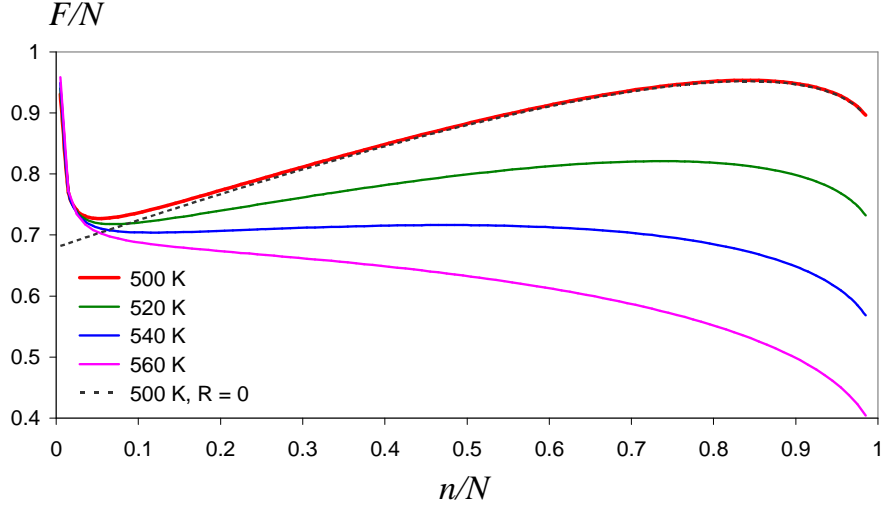

**Fig. 6.** Typical dependencies of free energy (22) on *n* for temperatures 500K (red line), 520K (green line), 540K (blue line) and 560K (pink line). Dependence for the free energy on *n* for *R*=0 and temperature 500K is plotted by black dash line. Here we used Gaussian approximation for the entropy of SCS and the parameters: $T_0$=600K, $E_0$=5kJ/mol, $N$=5000, $m$=40, $E_1$=$E_2$=$E_0$/3, $R=N^{1/2}a$.

The minimization of the free energy over *n* results in the following equation:

$$\frac{T_0 - T}{T_0} - \frac{2E_1}{mE_0} - \frac{2E_2}{E_0}\sqrt{\frac{m}{N-n}} - \frac{3kT}{2E_0}\left(\frac{R}{na}\right)^2 = 0 \qquad (23)$$

Numerical solution of equation (23) as a function of temperature is presented in Fig.7. In the limit *n*<<*N*, equation (23) can be expanded in the series over the small parameter *n/N* and solved analytically:

$$n_{eq}(T,R) = n_{0,eq}\left(1 + \frac{n_{0,eq}}{4N}\frac{T_1 - T_0^*}{T_1 - T} + O(N^{-2})\right) \qquad (24)$$

where

$$n_{0,eq}(T,R) = \frac{R}{a}\sqrt{\frac{3kT_0}{2E_0}\frac{T}{(T_1 - T)}} \qquad (25)$$

To simplify notations, here we introduced the melting temperature $T_0^*$ that corresponds to the melting temperature of an imaginary state where each chain forms a single nucleus. In this stage *R*=0, *i.e.* the second term in the right hand side of equation (22) is equal to zero:

$$T_0^* = T_0\left(1 - \left(\frac{2E_1}{mE_0} + \frac{4E_2}{E_0}\sqrt{\frac{m}{N}}\right)\right) \qquad (26)$$

The melting temperature, $T_0^*$, is shifted from $T_0$ by the surface energy of the nuclei.

The singularity temperature $T_1$ is slightly higher than $T_0^*$:

$$T_1 = T_0^* + T_0 \frac{2E_2}{E_0} \sqrt{\frac{m}{N}} \qquad (27)$$

The optimal number of the monomers in the SCS, $n_{0,eq}$, calculated from equation (25) is presented in Fig.7 as a function of temperature. As follows from Fig.7 the approximate analytical solution of equation (22), $n_{0,eq}$, substantially differs from the exact numerical solution only in small temperature interval near $T_1$, where the local minimum represents the metastable state. At the temperatures lower than $T_0^*$ the second term in the brackets in the right hand side of equation (24) is small. The leading term $n_{0,eq}$ calculated from equation (25) has the same functional dependence of the model parameters as $n_{eq}$ calculated from equation (9). The influence of the surface energy of the nuclei results in the shift of the singularity temperature from $T_0$ to $T_1$. In the range, where the nuclei are stable ($T < T_0^*$) the impact of surface energy on $n_{eq}$ is negligible.

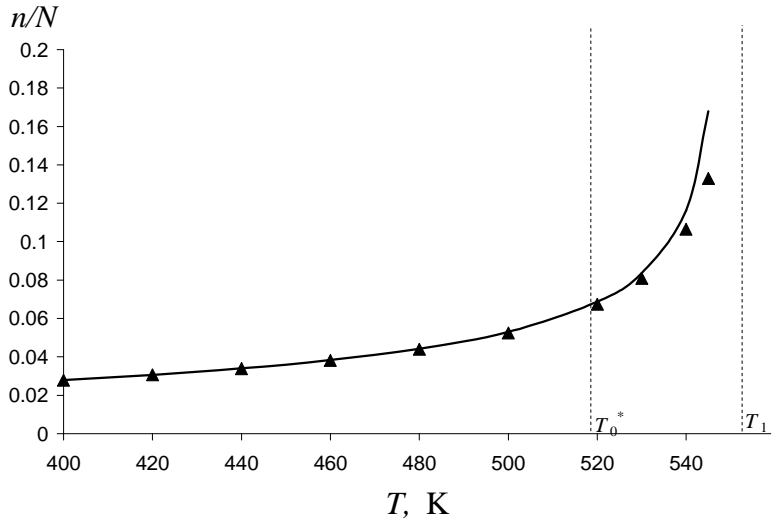

**Fig. 7.** Exact solution of equation (22) (solid line) and approximate solution, $n_{0,eq}$, (triangles) calculated by equation (25). Here we used Gaussian approximation for the entropy of SCS and the parameters: $T_0=600K$, $E_0=5kJ/mol$, $N=5000$, $m=40$, $E_1=E_2=E_0/3$, $R=N^{1/2}a$.

Below we perform initial qualitative validation of the model using two stress-strain experimental curves available for PTFE in the temperature interval between $T_g$ and $T_0$ [39] presented in Fig. 1. The PTFE have a lamellar structure with the thickness of lamellas larger than 100 nm (the breadths can be even larger) [60]. Two lamellas are connected by multiple SCS. In the current model, we consider a fraction of the PTFE lamella between two SCS and one of this SCS as a repeating unit. We estimate the yield stress of the polymer above the glass transition temperature as

$$\sigma_{YS}(T) \approx f_{eq}(T)\frac{1}{L^2} \quad (28)$$

Here $L^2$ is a mean cross-sectional area per one SCS, which is used as a fitting parameter. In general, $L$ is a function of temperature. It depends on temperature through $n(T)$ according to equation (21) and (25). However, as shown in Fig.7, $n$ is much smaller than $N$ for temperature range $T < T_0^*$. Therefore, the dependence of $L(T)$ is weak and can be neglected. To obtain the explicit equation for $\sigma_{YS}$ we utilize the Gaussian approximation for the entropy of the SCS. Substituting equation (11) into (28) we obtain

$$\sigma_{YS} = \frac{E_0}{aL^2}\sqrt{\frac{6kT}{E_0}\left(\frac{T_1 - T}{T_0}\right)} \quad (29)$$

The current model is inapplicable below the glass transition temperature, so that the low-temperature limit is out of the scope of current model.

From the experimental value of the yield stress at 260°C we estimate $L^2$ and make sure that the obtained value is in the right range. Then using the obtained value of $L$ we predict $\sigma_{YS}$ at 204°C. The experimental stress of PTFE at temperature 260°C and 100% strain is approximately equal to 2.2 MPa. The melting temperature of PTFE is 327°C, specific heat of fusion is approximately 5 kJ per mole of $CF_2$ monomers [61] and the monomer size is approximately 1 Å. Using equation (19) and these parameters we calculate the mean size of the nuclei of PTFE of $L$ = 6.3 nm. Using equation (19) we calculate the yield stress at $T = 204$°C, $\sigma_{YS} = 2.8$ MPa. The experimental stress of PTFE at $T = 204$°C and 100% strain is approximately equal to 3.4 MPa, which falls inside the 20% accuracy level.

## 4. Atomistic modeling of polymer nuclei

In this section, we present the results of molecular dynamic (MD) simulation of polymer chains folding, nuclei formation during cooling, and the nuclei melting during heating. The molecular dynamic models of the polyethylene (PE) crystal containing one and two chains were implemented in LAMMPS software package [62]. The united atom version of the simplistic Dreiding force-field [63] was used to model the crystallization/melting of the PE nucleus. The Nose-Hoover style thermostat and (when needed) barostat were used in these calculations [64]. The results were used to support the hypothesis of the nucleation through formation of the folds at the early stage of crystallization of the polymer with flexible chains.

The MD modeling of polymer melt crystallization in the system containing long chains with sufficient number of entanglements would require the simulation time much longer than the time of reptation between two entanglements. However, molecular dynamic is typically capable of modeling too small time intervals, less than several decades of nanoseconds. To overcome this limitation we prepared the MD models of polymer melt containing one and two polymer molecules in periodic boundary conditions and studied the crystallization process in such systems.

The MD model of single PE crystal consisted of one chain was prepared using the following systematic procedure. A straight all-trans PE chain comprised of 1000 carbon atoms (500 monomeric units) in non-periodic boundary conditions was constructed. The chain was heated up to 1000 K with ends fixed in their initial positions. Then the distance between the chain ends was gradually decreased during 1 ns. The temperature was simultaneously decreased to 500 K. A final distance between the chain ends was chosen so that the system density was approximately equal to the density of the solid PE, $\rho = 0.9$ g/cm$^3$. As a result, the chain packed irregularly into a spherical drop, as shown in Fig. 8(a).

Then the chains ends were unfixed. After 1 ns at 500 K the chain formed several quasi-crystalline regions, as shown in Fig. 8(b). In each quasi-crystal the chain fragments were oriented parallel to each other. The overall chain shape became non-spherical. Subsequently, the system was subjected to thermalization at 500 K in NPT ensemble and periodic boundary conditions for 5 ns. After thermalization all chain fragments were oriented parallel to each other. The front view of the final crystal is shown in Fig. 8(c) and side view is shown in Fig. 8(d).

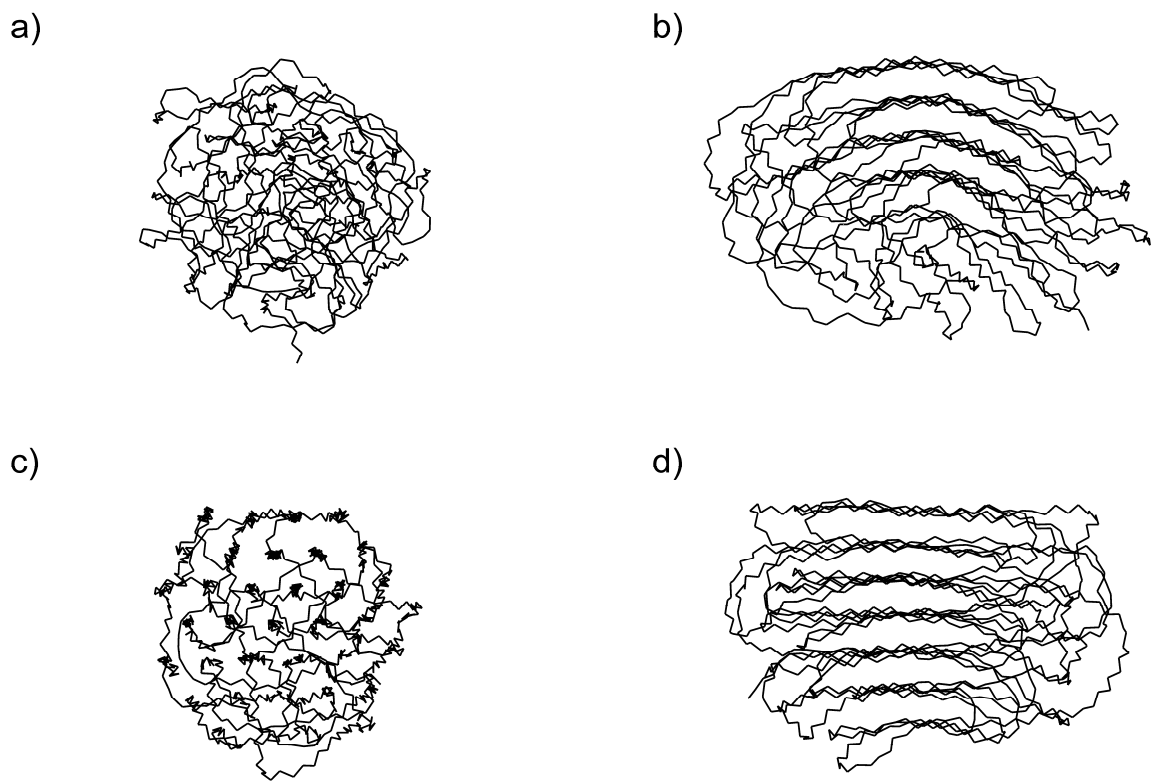

**Fig. 8.** Snapshots of one chain at different time moments during the cooling.

To study the peculiarities of the polymer melting the MD model of PE crystal containing one polymer chain was subjected to melting in the box with periodic boundary conditions. The temperature of the system was increased stepwise from 500 K up to 900 K in NPT ensemble. In each step, we raised the temperature by 50 K for 1 ns and maintained it constant for next 5 ns for thermalization of the system. This step was repeated until the temperature 900 K was reached. The snapshots of the system at four temperatures were shown in Fig. 9.

At the temperatures 500 K and 650 K the system is a crystal. At 650 K the perturbations of the regular structure of the crystal are obtained and the chain segments form loops at the edge of the crystal, as shown in Fig. 9(b). That indicates the temperature approaches the melting point. At 700 K the perturbations of the crystalline order increase and the crystal changes the shape and elongates, as shown in Fig. 9(c). This re-crystallization leads to decrease of the crystal energy, as shown in Fig. 10, and is a result of increase of monomers mobility in the crystal near the melting point. The decrease of the crystal energy is caused by the reduction of the number of energetically unfavorable edge monomers, see Fig. 9(a). The crystalline order disappears at the temperature 750 K, as shown in Fig. 9(d), and we refer to this temperature as the melting point. The experimental melting point of PE is about 410 K. We attribute this disagreement of simulated melting point with experimental one to roughness of Dreiding force field used in our MD simulation. Therefore, this simulation predicts only qualitative behavior of PE chain.

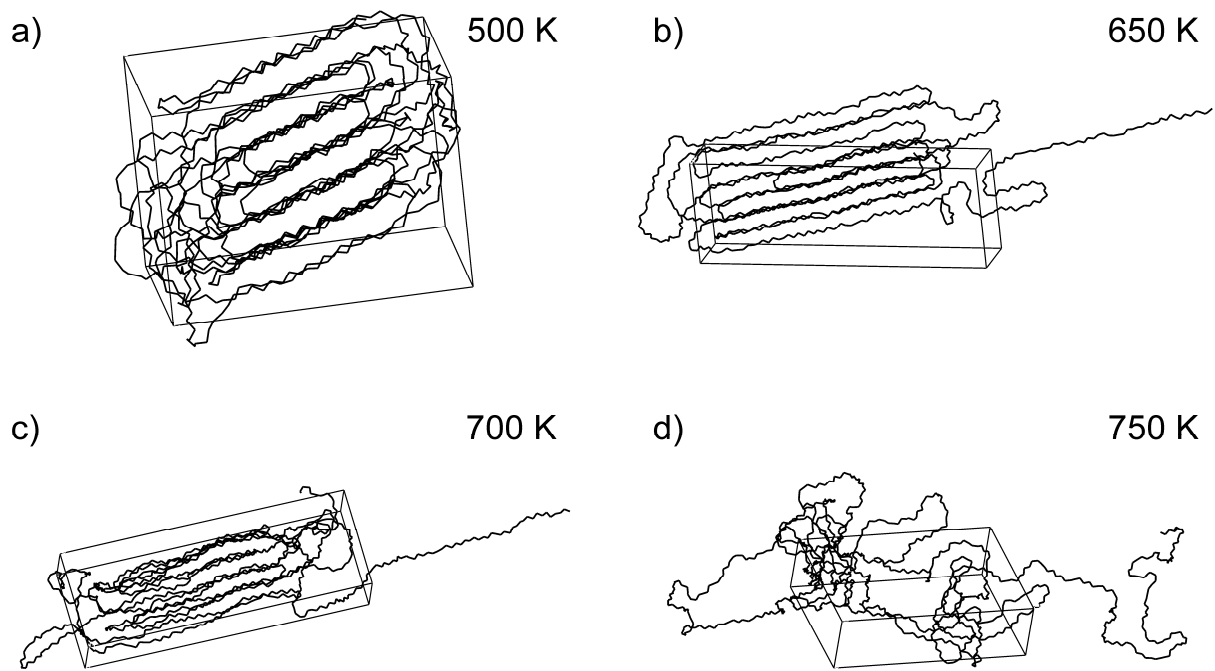

**Fig. 9.** Snapshots of one chain in periodic boundary conditions at different temperatures during the heating.

The potential energy of the system is plotted as a function of the temperature in Fig. 10. The potential energy gradually increases in the temperature regions between 500 K and 650 K, and above 750 K. The slight decrease of the potential energy is observed when the temperature is changed from 650 K to 700 K. That is caused by elongation of the crystal and reduction of the number of the edge monomers (see Fig. 9(a)), which are energetically unfavorable. Further increase of the temperature leads to the melting of the crystal and the stepwise increase of the potential energy of the system. The disappearance of crystalline order at $T = 750$ K can also be seen in the snapshot of the system at $T = 750$ K, see Fig. 9(d).

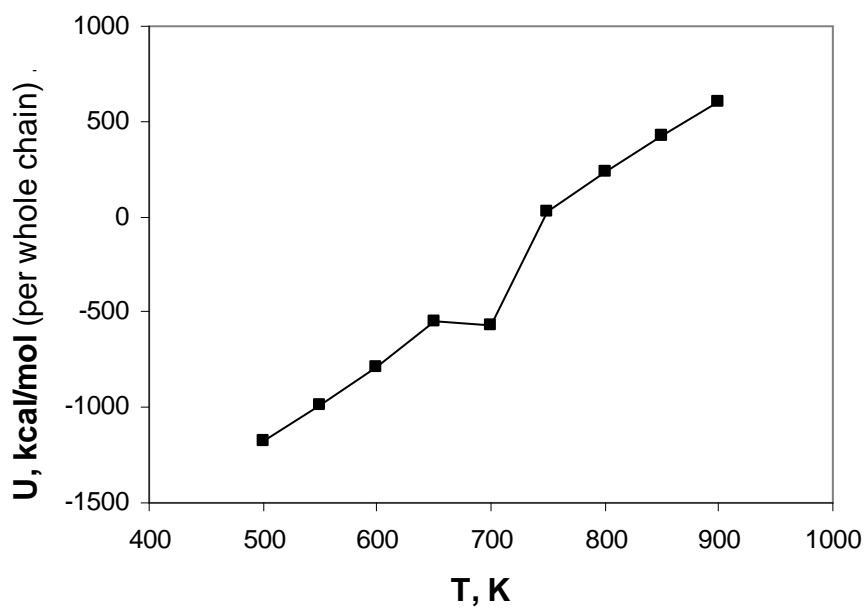

**Fig. 10.** Dependence of the potential energy of the one-chain crystal on the temperature.

To understand the topology of multi-chain PE crystal we prepared the MD model of the crystal containing two polymer chains as follows. Two chains with 1000 monomers in one chain were placed into a large box at temperature 800 K. The MD simulation was started and the box size was gradually decreased until the system reached the density approximately equal to the density of the solid PE. Then the temperature was gradually decreased from 800 K to 200 K. At the temperature approximately equal to 550 K the crystallization of the system was observed. Crystallization manifests itself through the ordering of the system. The snapshot of two-chain crystal is shown in Fig. 11. The crystal looks like a single structure that contains folded segments of each chain. Apparently, there is no segregation of the chains, which would manifest itself as separate crystals or structural defects inside one crystal. To distinguish two chains in Fig. 11 we marked one chain by black color and the other one by grey. It is seen from Fig. 11 that the chains are mixed inside the crystal without disrupting of the crystal structure.

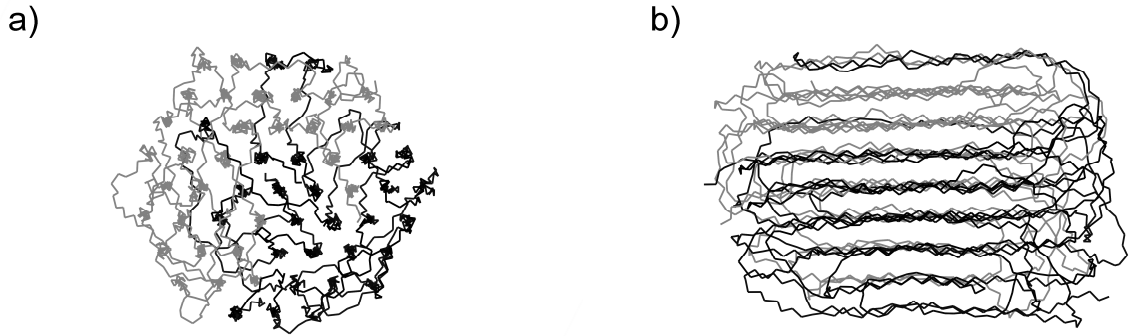

**Fig. 11.** Front (a) and side view (b) of the two-chain crystal. One chain is indicated by grey color and other chain is indicated by black color.

## 5. Discussion and conclusions

The microscopic model of semi-crystalline polymer with flexible chains is presented in this paper. The model is based on the assumption that, below the melting temperature, $T_0$, the semi-crystalline polymer comprises of crystal nuclei and randomly packed chain segments that connect the nuclei. At the melting temperature, a large number of crystal nuclei are formed along the polymer chain and grow during cooling. Topological restrictions caused by the connectivity of the monomers into the chain impede crystallization process in the polymer and the major fraction of the nuclei does not grow into the large crystals even at the temperatures far below $T_0$. These nuclei are connected by the randomly packed stretched chain segments (SCS). The nuclei of the polymers with flexible chains (PE, PTFE) are the folds comprising of the segments of one or more chains. The folded structure of the nuclei results in large reversible deformability of the polymers. Unfolding of the nuclei under applied external stress leads to large polymer deformation without loss of integrity. Subsequent folding of the chains after unloading results in reversibility of the large deformation, i.e. the polymer recovers its initial shape.

We assume that the fast exchange of the monomers between the nuclei and the SCS leads to thermodynamic equilibrium between the nuclei and the SCS above $T_g$. The simple thermodynamic model predicts that the equilibrium size of the nuclei and the SCS depend on the temperature. The number of the monomers in the nuclei decreases with increase of the temperature. The model also predicts that the SCS are under tension. This tension is driven by the tendency of the polymer to crystallize. The increase of the temperature causes decrease of the SCS tensile force. At the melting temperature, $T_0$, the tensile force approach zero. We assume that the SCS tensile force governs the elastic properties of the polymer in the temperature interval between $T_g$ and $T_0$.

A typical experimental stress-strain curve of high-elastic PTFE is presented in Fig. 1. It demonstrates fast linear increase of the stress at $\varepsilon<30\%$, which is referred to in the literature as a Hookean regime. The plateau in the stress-strain curve of PTFE follows the Hookean regime. This plateau resembles the yield stress in the polymer below $T_g$. However, the large deformation of high-elastic polymer is reversible while the large deformation below $T_g$ is irreversible. We assume that the yield stress of the high-elastic polymer is governed by the SCS tension force. An external force exceeding the total tensile force between the polymer folds is required for large deformation. The simple 1D thermodynamic model developed in this paper predicts that the stress-strain curve of high-elastic polymer possesses the flat region, which qualitatively agrees with the experimental stress-strain curve of PTFE, shown in Fig. 1.

To obtain a closed form of analytical expression for the SCS tensile force, the Gaussian chain approximation for the conformational entropy of the SCS was utilized. The simplest approximation predicts that the size of the nuclei decreases as $\sqrt{T_0 - T}$ when the temperature approaches $T_0$. Also, the model predicts that the tensile force of the SCS and yield stress of high-elastic polymer tend to zero as $\sqrt{T_0 - T}$, when the temperature approaches $T_0$. To improve the model predictions close to the melting point the surface energy of the nuclei was taken into account. The influence of the surface energy results in termination of the gradual decrease of the nuclei size upon approaching the melting point. At the temperature $T_0^*$ the nuclei become thermodynamically unstable and abruptly melt.

The more general expression for the entropy of the SCS was used assuming that the entropy of the SCS depends only on the scaling parameter $x = \dfrac{na}{R}$, where $n$ is a number of the monomers in the SCS, $a$ is the monomer size and $R$ is a distance between the ends of SCS (see Fig. 2). In both cases, for the Gaussian and for the general expression for the entropy, the model predicts that the tensile force depends only on the temperature and does not depend on the distance $R$ between the nuclei. As a consequence, the model predicts that the stress in the polymer is independent of the strain when an external force is higher then the threshold value. This threshold value manifests itself as the apparent yield stress in the stress-strain curve of the high-elastic polymer.

The temperature dependence of the PTFE yield stress above $T_g$ is calculated using the mean cross-section area per one SCS as a fitting parameter. The model is in a satisfactory agreement (~20% accuracy) with the experiment. This qualitative validation of the model through the limited number of the experimental data of course cannot be considered as a complete and rigorous validation.

To verify the model assumptions, the crystallization and melting processes of the single PE crystal consisted of one and two polymer chains were modeled by molecular dynamics. The MD modeling shows that the PE chains form the folds under crystallization. Moreover, two chains form a single crystal. No chains segregation was observed inside the crystal, i.e. the chains mix inside the crystal without disrupting of the crystal structure. That indicates that the nucleus of the flexible polymer consists of the segments of several chains, and it is connected with several other nuclei by the SCS. That results in formation of a complex network in the amorphous area of the polymer, which consists of the nuclei with multiple connections by the SCS. The modeling of the melting process shows that the perturbations of the regular structure of the crystal appear at temperatures below the melting point of about $T_0/5$. These perturbations of the crystal structure start from formations the loops at the edge of the crystal and develop through the internal defects of the crystal regular shape, as shown in Fig. 9(b).

The results obtained in this work are very similar to fundamental conclusions of Di Marzio and Guttman paper [39]. In [39] the peeling force of a single polymer chain absorbed on the surface was calculated as a function of the temperature and the distance between the chain end and the surface. In both cases, the underlying physical phenomenon is the equilibrium between the low and high energy states of the chain monomers. For high-elastic deformation, this is the equilibrium between the nucleus and the SCS. For chain peeling modeled in [39], this is the equilibrium between the absorbed and peeled out fractions of the chain. As a consequence, both the peeling force and the tensile force have similar qualitative features. To peel the polymer chain from the surface the peeling force should exceed a threshold value that is a function of the temperature, the absorption energy per one monomer and the monomer size. The critical value of tensile force, which governs the apparent yield stress in high-elastic polymer calculated in the present paper, is also a function of the temperature, specific melting heat and the monomer size. Both the peeling force and tensile force are independent of the displacement when the force is higher then the threshold. The peeling force is independent of the distance between the chain end and the surface and the tensile force is independent of the length of the SCS between two nuclei. Moreover, both models utilize the Gaussian approximation for the entropy of the chain in the high-energy state. Consequently, both models predict the square root dependence, $\sqrt{T_0 - T}$, for the force as a function of the deviation of the function from the critical value, $T_0$.

**Acknowledgement**

The authors gratefully acknowledge Dr. M. McQuade and Dr. D. Parekh of UTC for the interest to the work, inspiring discussion and support, and Professor J.M. Deutch of Massachusetts Institute of Technology and Professor G.M. Whitesides of Harvard University and Dr. Charles Watson of Pratt & Whitney for interesting discussion of the results and suggestions for future work and potential applications.

**Appendix**

The entropy of the SCS in Gaussian approximation is calculated as follows:

$$S(n,R) = k \ln(Q(n,R)) \tag{A1}$$

Here $Q(n,R)$ is the number of configurations of random walk containing $n$ steps with distance $R$ between initial and finish points. The number of state of the random walk reads:

$$Q(n,R) = z^n \frac{1}{(2\pi n)^{3/2}} \exp\left(-\frac{3R^2}{2na^2}\right) \tag{A2}$$

Here $z$ is the number of configurations per one step of the random walk and $a$ is the length of the step. Substituting equation (A2) into (A1) we obtain:

$$S(n,R) = nk \ln(z) - k\frac{3R^2}{2na^2} - k\frac{3}{2}\ln(2\pi n) \tag{A3}$$

Taking advantage of equation (5) and substituting equation (A3) into (1) we obtain the equation for the free energy of repeating element:

$$F(n,R) = nE_0 \frac{T_0 - T}{T_0} - \frac{3kT}{2}\left(\frac{R^2}{na^2} + \ln(2\pi n)\right) \tag{A4}$$

Differentiating equation (A4) with respect to $n$ we obtain equation for the optimal number of the monomers:

$$E_0 \frac{T_0 - T}{T_0} + \frac{3kT}{2}\left(\frac{R}{na}\right)^2 - \frac{3kT}{2}\frac{1}{n} = 0 \tag{A5}$$

Solving equation (A5) with respect to $n$ we obtain the optimal number of the monomers in the SCS:

$$n_{eq} = \frac{1}{2}\frac{3kT_0}{2E_0}\frac{T}{(T_0 - T)}\left[\sqrt{1 + 4\frac{2E_0}{3kT_0}\frac{(T_0 - T)}{T}\left(\frac{R}{a}\right)^2} - 1\right] \tag{A6}$$

In the vicinity of the melting point, when $4\frac{2E_0}{3kT_0}\frac{(T_0 - T)}{T} \ll \left(\frac{a}{R}\right)^2$, the square root in the right hand side of equation (A6) can be expanded into the series and at $T=T_0$

$$n_{eq}(T_0) = \left(\frac{R}{a}\right)^2 \tag{A7}$$

In the melt, where the chain obeys a Gaussian statistics, $R = \sqrt{N}a$ and equations (A7) reads:

$$n_{eq}(T_0) = N \tag{A8}$$

Below the melting point, if $4\dfrac{2E_0}{3kT_0}\dfrac{(T_0 - T)}{T} \gg \left(\dfrac{a}{R}\right)^2$, the approximate equation for $n_{eq}$ reads:

$$n_{eq} \approx \frac{R}{a}\sqrt{\frac{2E_0}{3kT_0}\frac{(T_0 - T)}{T}} \tag{A9}$$

Equation (A9) coincides with equation (9).

**References**

[1] Huang X, Solasi R, Zou Y, Feshler M, Reifsnider K, Condit D, Burlatsky S and Madden T, *Mechanical endurance of polymer electrolyte membrane and PEM fuel cell durability*, 2006 *J. Polym. Sci., Part B: Polym. Phys.* **44** 2346

[2] Burlatsky S F, Gummalla M, O'Neill J, Atrazhev V V, Varyukhin A N, Dmitriev D V and Erikhman N S, *A mathematical model for predicting the life of PEM fuel cell membranes subjected to hydration cycling*, 2012 *J. Power Sources* **215** 135

[3] Regel' V R, Slutsker A I and Tomashevskiĭ É E, *The kinetic nature of the strength of solids*, 1972 *Sov. Phys. Uspekhi* **15** 45

[4] Oshanin G, Moreau M and Burlatsky S, *Models of chemical reactions with participation of polymers*, 1994 *Adv. Colloid Interface Sci.* **49** 1

[5] Ehrenstein G W, 2001 *Polymeric materials: Structure, properties, applications* (Munich: Hanser)

[6] Cheng S Z D and Keller A, *The role of metastable states in polymer phase transitions: Concepts, principles, and experimental observations*, 1998 *Ann. Rev. Mater. Sci.* **28** 533–562

[7] Di Lorenzo M L, Righetti M C, Cocca M and Wunderlich B, *Coupling between Crystal Melting and Rigid Amorphous Fraction Mobilization in Poly(ethylene terephthalate)*, 2010 *Macromolecules* **43** 7689

[8] Ward I M, 1983 *Mechanical properties of Solid Polymers* (New York: Wiley)

[9] Sweeney I M and Ward J, 2004 *An introduction to the mechanical properties of solid polymers* (New York: Wiley)

[10] Berthier L and Biroli G, *Theoretical perspective on the glass transition and amorphous materials*, 2011 *Rev. Mod. Phys.* **83** 587

[11] Khanna Y P, Turi E A, Taylor T J, Vickroy V V and Abbott R F, *Dynamic mechanical relaxations in polyethylene*, 1985 *Macromolecules* **18** 1302

[12]   Oshima A, Ikeda S, Seguchi T and Tabata Y, *Change of molecular motion of polytetrafluoroethylene (PTFE) by radiation induced crosslinking*, 1997 *Radiat. Phys. Chem.* **49** 581

[13]   Oleynik E, *Plastic deformation and mobility in glassy polymers*, 1989 *Relaxation in Polymers* Progress in Colloid and Polymer Science vol 80, ed M Pietralla and W Pechhold (Berlin, Heidelberg: Springer) pp 140–50

[14]   Alves N M, Mano J F and Gómez Ribelles J L, *Structural relaxation in a polyester thermoset as seen by thermally stimulated recovery*, 2001 *Polymer* **42** 4173

[15]   Alves N M, Mano J F and Gómez Ribelles J L, *Molecular mobility in polymers studied with thermally stimulated recovery. II. Study of the glass transition of a semicrystalline PET and comparison with DSC and DMA results*, 2002 *Polymer* **43** 3627

[16]   Alves N M, Mano J F and Gómez Ribelles J L, *Molecular mobility in polymers studied with thermally stimulated recovery - I. Experimental procedures and data treatment*, 2002 *J. Therm. Anal. Calorim.* **70** 633

[17]   Flory P J, 1971 *Principles of polymer chemistry* (Ithaca: Cornell University Press)

[18]   Huggins M L, *Solutions of long chain compounds*, 1941 *J. Chem. Phys.* **9** 440

[19]   Huggins M L, *Thermodynamic properties of solutions of long-chain compounds*, 1942 *Ann. NY Acad. Sci.* **43** 1

[20]   Doi M and Edwards S F, 1986 *The theory of polymer dynamics* (Oxford: Clarendon Press)

[21]   de Gennes P G, *Reptation of a polymer chain in the presence of fixed obstacles*, 1971 *J. Chem. Phys.* **55** 572

[22]   de Gennes P J, 1986 *Scaling concepts in polymer physics* (Oxford: Cornell University Press)

[23]   Flory P J, *The configuration of real polymer chains*, 1949 *J. Chem. Phys.* **17** 303

[24]   Lifshits I M, Grosberg A Y and Khokhlov A R, *Volume interactions in the statistical physics of a polymer macromolecule*, 1979 *Sov. Phys. Uspekhi* **22** 123

[25]   Grosberg A Y, Nechaev S K, Shakhnovich E I, *The role of topological constraints in the kinetics of collapse of macromolecules*, 1988 *Journal de physique* **49** 2095

[26]   Grosberg A Y and Nechaev S K, *Topological constraints in polymer network strong collapse*, 1991 *Macromolecules* **24** 2789

[27]   Rubinstein M and Colby R H, 2003 *Polymer physics* (Ithaca: Oxford University Press)

[28]   Grosberg A Y and Khokhlov A R, 1994 *Statistical physics of macromolecules* (New York: AIP Press)

[29]   James H M and Guth E, *Theory of the elastic properties of rubber*, 1943 *J. Chem. Phys.* **11** 455

[30]   Flory P J, *Network structure and the elastic properties of vulcanized rubber.*, 1944 *Chem. Rev.* **35** 51

[31]   Treloar L R G, 1975 *The physics of rubber elasticity* (Oxford: Clarendon Press)

[32]   Flory P J, *Theory of elasticity of polymer networks. The effect of local constraints on junctions*, 1977 *The Journal of Chemical Physics* **66** 5720

[33]   Edwards S F, *The statistical mechanics of polymerized material*, 1967 *Proceedings of the Physical Society* **92** 9

[34]   Heinrich G, Straube E and Helmis G, *Rubber elasticity of polymer networks: Theories*, 1988 *Polymer Physics* vol 85 (Berlin, Heidelberg: Springer Berlin Heidelberg) pp 33–87

[35]   Kovac J, *Modified Gaussian model for rubber elasticity*, 1978 *Macromolecules* **11** 362

[36]   Luo W and Tan J, *A hybrid hyperelastic constitutive model of rubber materials*, 2008 *Advances in Heterogeneous Material Mechanics* 2nd International Conference on Heterogeneous Material Mechanics (Huangshan, China)

[37]   Dobrynin A V, Carrillo J-M Y and Rubinstein M, *Chains Are More Flexible Under Tension*, 2010 *Macromolecules* **43** 9181

[38]   Bergström J S and Hilbert L B, *A constitutive model for predicting the large deformation thermomechanical behavior of fluoropolymers*, 2005 *Mech. Mater.* **37** 899

[39]   *DuPont Teflon PTFE fluoropolymer resin: Properties Handbook* [Online] http://www2.dupont.com/Teflon_Industrial/en_US/assets/downloads/h96518.pdf

[40]   Haward R N, *Strain hardening of thermoplastics*, 1993 *Macromolecules* **26** 5860

[41]   Haward R N, *The derivation of a strain hardening modulus from true stress-strain curves for thermoplastics*, 1994 *Polymer* **35** 3858

[42]   Haward R N, *The application of a Gauss-Eyring model to predict the behavior of thermoplastics in tensile experiments*, 1995 *J. Polym. Sci., Part B: Polym. Phys.* **33** 1481

[43]   Eyring H, *Viscosity, plasticity, and diffusion as examples of absolute reaction rates*, 1936 *J. Chem. Phys.* **4** 283

[44]   van Dommelen J, Parks D, Boyce M, Brekelmans W and Baaijdens F, *Micromechanical modeling of the elasto-viscoplastic behavior of semi-crystalline polymers*, 2003 *J. Mech. Phys. Solids* **51** 519

[45]   Hoffman J D, *Regime III crystallization in melt-crystallized polymers: The variable cluster model of chain folding*, 1983 *Polymer* **24** 3

[46]   Kozlov G V and Novikov V U, *A cluster model for the polymer amorphous state*, 2001 *Phys.-Usp.* **44** 681

[47]   Liu C and Muthukumar M, *Langevin dynamics simulations of early-stage polymer nucleation and crystallization*, 1998 *The Journal of Chemical Physics* **109** 2536

[48]   Muthukumar M and Welch P, *Modeling polymer crystallization from solutions*, 2000 *Polymer* **41** 8833

[49] Muthukumar M, *Modeling Polymer Crystallization*, 2005 *Interphases and Mesophases in Polymer Crystallization III* Advances in Polymer Science vol 191, ed G Allegra (Springer Berlin / Heidelberg) pp 241–74

[50] Larini L, Barbieri A, Prevosto D, Rolla P A and Leporini D, *Equilibrated polyethylene single-molecule crystals: molecular-dynamics simulations and analytic model of the global minimum of the free-energy landscape*, 2005 *Journal of Physics: Condensed Matter* **17** L199

[51] Queyroy S and Monasse B, *Effect of the molecular structure of semicrystalline polyethylene on mechanical properties studied by molecular dynamics J. Appl. Polym. Sci.* **125** 4358

[52] Di Marzio E A and Guttman C M, *Peeling a polymer from a surface or from a line*, 1991 *J. Chem. Phys.* **95** 1189

[53] Muthukumar M, *Shifting Paradigms in Polymer Crystallization*, 2007 *Progress in Understanding of Polymer Crystallization* Lecture Notes in Physics vol 714, ed G Reiter and G Strobl (Berlin / Heidelberg: Springer) pp 1–18

[54] Grosberg A and Nechaev S, *Polymer topology*, 1993 *Polymer Characteristics* Advances in Polymer Science vol 106 (Berlin, Heidelberg: Springer) pp 1–29

[55] Rouse P E, *A theory of the linear viscoelastic properties of dilute solutions of coiling polymers*, 1953 *J. Chem. Phys.* **21** 1272

[56] Li L, Chan C-M, Li J-X, Ng K-M, Yeung K-L and Weng L-T, *A direct observation of the formation of nuclei and the development of lamellae in polymer spherulites*, 1999 *Macromolecules* **32** 8240

[57] Nozue Y, Kurita R, Hirano S, Kawasaki N, Ueno S, Iida A, Nishi T and Amemiya Y, *Spatial distribution of lamella structure in PCL/PVB band spherulite investigated with microbeam small- and wide-angle X-ray scattering*, 2003 *Polymer* **44** 6397

[58] Flory P J and Yoon D Y, *Molecular morphology in semicrystalline polymers*, 1978 *Nature* **272** 226

[59] Sommer J-U, *Theoretical Aspects of the Equilibrium State of Chain Crystals*, 2007 *Progress in Understanding of Polymer Crystallization* Lecture Notes in Physics vol 714, ed G Reiter and G Strobl (Berlin / Heidelberg: Springer)

[60] Tonelli A E, *Polyethylene and polytetrafluoroethylene crystals: chain folding, entropy of fusion and lamellar thickness*, 1976 *Polymer* **17** 695

[61] Starkweather Jr H W, Zoller P, Jones G A and Vega A J, *The heat of fusion of polytetrafluoroethylene*, 1982 *J. Polym. Sci.: Polymer Physics Edition* **20** 751

[62] Plimpton S, *Fast parallel algorithms for short-range molecular dynamics*, 1995 *Journal of Computational Physics* **117** 1

[63] Mayo S L, Olafson B D and Goddard W A, *DREIDING: a generic force field for molecular simulations*, 1990 *J. Phys. Chem.* **94** 8897

[64] Shinoda W, Shiga M and Mikami M, *Rapid estimation of elastic constants by molecular dynamics simulation under constant stress*, 2004 *Phys. Rev. B* **69** 134103

**Figure titles**

**Fig. 1.** Stress-strain curves of rubber at room temperature [23] and PTFE in high-elastic state at T=204°C and T=260°C [29].

**Fig. 2.** The evolution of polymer structure at fixed temperature below $T_0$. a) polymer melt, b) beginning of nucleation process (nuclei A and B are the neighboring nuclei belonging to one chain), c) termination of nuclei growth (nuclei A and B are arrested by entanglements), the SCS is stretched between the nuclei A and B.

**Fig. 3.** The polymer structure at different temperatures a) $T = 0$, the number of the monomers in the SCS is minimal and the SCS is fully stretched, b) $T > 0$, some fraction of the monomers transfers from the nuclei into the SCS, c) $T \to T_0$ (melting temperature), the major fraction of the monomers transfers into the SCS.

**Fig. 4.** The unfolding of the nuclei induced by large deformation. The system a) before deformation, b) after deformation. The distance $R$ and the number of the monomers in the SCS after deformation are larger than that before the deformation. The number of the monomers in the nuclei is smaller after deformation than that before the deformation.

**Fig. 5.** The sketch of folded nucleus containing $(N-n)$ monomers.

**Fig. 6.** Typical dependencies of free energy (22) on $n$ for temperatures 500K (red line), 520K (green line), 540K (blue line) and 560K (pink line). Dependence for the free energy on $n$ for $R=0$ and temperature 500K is plotted by black dash line. Here we used Gaussian approximation for the entropy of SCS and the parameters: $T_0=600K$, $E_0=5kJ/mol$, $N=5000$, $m=40$, $E_1=E_2=E_0/3$, $R=N^{1/2}a$.

**Fig. 7.** Exact solution of equation (22) (solid line) and approximate solution, $n_{0,eq}$, (triangles) calculated by equation (25). Here we used Gaussian approximation for the entropy of SCS and the parameters: $T_0=600K$, $E_0=5kJ/mol$, $N=5000$, $m=40$, $E_1=E_2=E_0/3$, $R=N^{1/2}a$.

**Fig. 8.** Snapshots of one chain at different time moments during the cooling.

**Fig. 9.** Snapshots of one chain in periodic boundary conditions at different temperatures during the heating.

**Fig. 10.** Dependence of the potential energy of the one-chain crystal on the temperature.

**Fig. 11.** Front (a) and side view (b) of the two-chain crystal. One chain is indicated by grey color and other chain is indicated by black color.